\documentclass[12pt]{article}

\usepackage{epsfig}
\usepackage{graphicx}
\usepackage{natbib}
\usepackage[usenames,dvipsnames]{color}
\usepackage{amssymb}
\usepackage{amsmath} 
\usepackage{mathrsfs}
\usepackage[colorlinks,citecolor=blue,urlcolor=blue]{hyperref}

\usepackage{setspace}
\usepackage[font = {it}]{caption}

\RequirePackage{bm}

\allowdisplaybreaks  %allow equations break

\title{A Nonparametric Method for Producing Isolines of Bivariate Exceedance Probabilities}

\author
  {Daniel Cooley$^1$, 
  Emeric Thibaud$^2$\\ 
  Federico Castillo$^3$,
  Michael F. Wehner$^4$\\
  $^1$Department of Statistics, Colorado State University\\
  $^2$Institute of Mathematics, Ecole Polytechnique F\'ed\'erale de Lausanne\\
  $^3$Department of Environmental Science, Policy and Management,\\
  University of California, Berkeley\\
  $^4$Lawrence Berkeley National Laboratory\\
  }
\date{\today}

\begin{document}
{\singlespacing
\maketitle
}%ends singlespacing

%%%%%%%%%%%%%%%%%%%%%%%%%%%%%%%%%%%%%%%%%%%%%%%%%%
%%%%%%%%%%%%%%%%%%%%%%%%%%%%%%%%%%%%%%%%%%%%%%%%%%
%%%%%%%%%%%%%%%%%%%%%%%%%%%%%%%%%%%%%%%%%%%%%%%%%%
\begin{abstract}

We present a method for drawing isolines indicating regions of equal joint exceedance probability for bivariate data.
The method relies on bivariate regular variation, a dependence framework widely used for extremes. 
This framework enables drawing isolines corresponding to very low exceedance probabilities and these lines may lie beyond the range of the data.
The method we utilize for characterizing dependence in the tail is largely nonparametric.
Furthermore, we extend this method to the case of asymptotic independence and propose a procedure which smooths the transition from asymptotic independence in the interior to the first-order behavior on the axes.
We propose a diagnostic plot for assessing isoline estimate and choice of smoothing, and a bootstrap procedure to visually assess uncertainty.

\end{abstract}


%%%%%%%%%%%%%%%%%%%%%%%%%%%%%%%%%%%%%%%%%%%%%%%%%%%
%%NOTATION
%%%%%%%%%%%%%%%%%%%%%%%%%%%%%%%%%%%%%%%%%%%%%%%%%%%
%
%\section*{Notation}
%$\bm X$ random vector in original (untransformed) space\\
%$\bm Z$ random vector after transformation.

\doublespacing

%%%%%%%%%%%%%%%%%%%%%%%%%%%%%%%%%%%%%%%%%%%%%%%%%%
%INTRODUCTION
%%%%%%%%%%%%%%%%%%%%%%%%%%%%%%%%%%%%%%%%%%%%%%%%%%
\section{Introduction}

%goal 
We develop a tool which will draw isolines to indicate regions of equal joint exceedance probability for bivariate data.
By displaying these regions of low probability, researchers can visually assess probabilistic risk of rare bivariate extreme events.  
Importantly, impactful events can arise when the combination of variables is rare even if the individual variates are not at their highest values.
We employ results from multivariate extreme value (EV) theory which provide a framework for characterizing dependence in the tail of the distribution.
Although our method is largely nonparametric, we are able to extrapolate to describe events more extreme than any observed in the data record.

%motivating data
In Figure \ref{fig: motivation} we present two motivating data sets which we will examine in this work.
Details about the data are given in Sections \ref{sec: asyDepCase} and \ref{sec: asyIndepCase}.
The left panel shows data related to a southern California weather regime known as the Santa Ana winds, a windy and dry weather regime conducive for wildfires.
The points labeled ``C" and ``W" correspond to the ignition days of the Cedar and Witch Fires respectively, both of which were among the most destructive Santa Ana driven wildfires on record.
The right panel shows daily temperature measurements and relative humidity measurements for Karachi, Pakistan.
Here, risk is in terms of human health impacts which worsen by simultaneous hot and humid conditions.
Shown in black are six successive days in June 2015 which correspond to a heat wave which is blamed for the deaths of more than 700 people \citep{masood2015}.

These two examples illustrate why a bivariate extremes approach is necessary. 
The standard practice to diagnose the statistics of fires or heat waves would be to consider univariate statistics of some appropriate combined measure of the relevant individual variables such as a burn index or a human health index. 
However, as is evident by the spread of both sets of points in Figure \ref{fig: motivation}, the relationship between the two meteorological variables is complex which a combined variable cannot capture and potentially valuable information is lost.
Instead of relying on the aforementioned indices, an understanding of bivariate extreme behavior could improve response to the crisis by allocating resources in a more efficient manner.  
To aid in understanding bivariate extreme behavior, we would like to draw lines to indicate how frequently events this extreme, or even more extreme, can be expected to occur.

\begin{figure}[t]
  \begin{center}
    \includegraphics[scale = .75]{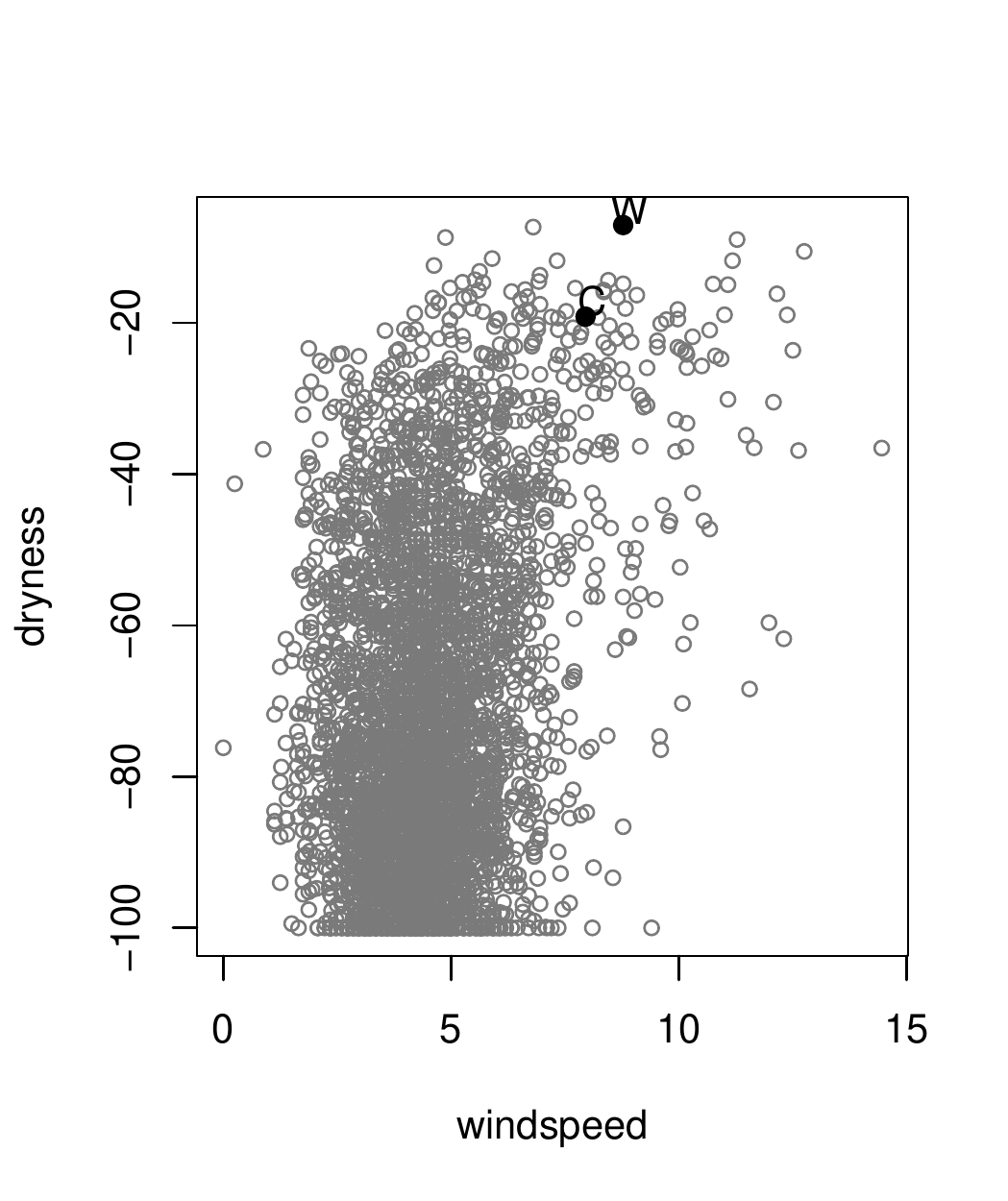}
    \includegraphics[scale = .75]{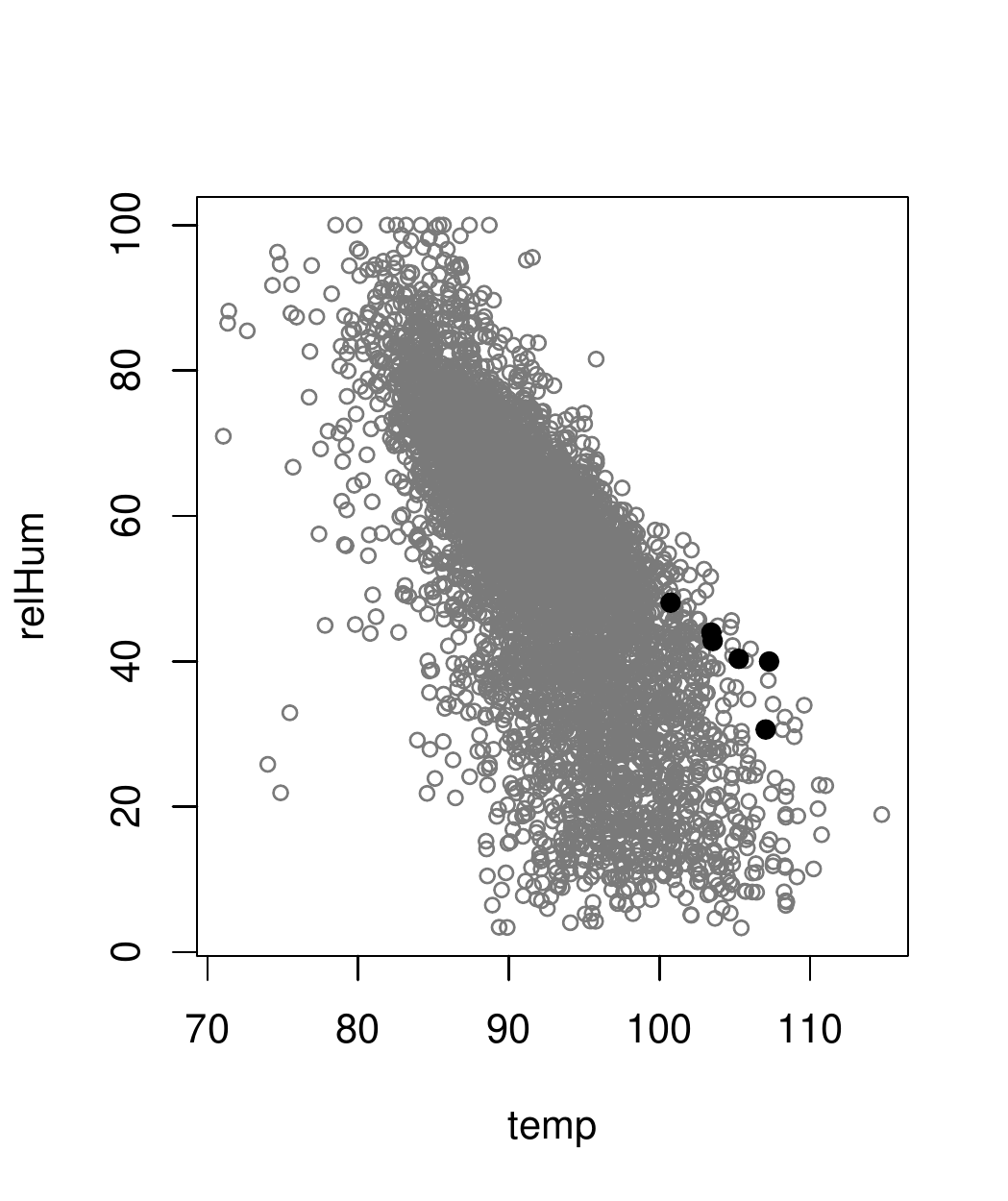}
  \end{center}
  \caption{Left panel:  windspeed and dryness from the Santa Ana dataset.  Point labeled ``C" corresponds to the date of the Cedar Fire, and point labeled ``W" corresponds to that of the Witch Fire.  Right panel:  temperature and relative humidity from the Karachi data set.  Dark solid circles correspond to the dates 06/18/2015-06/23/2015.}
  \label{fig:  motivation}
\end{figure}

%challenge in bivariate space as opposed to univariate
In the univariate case there is a one-to-one correspondence between probabilities and exceedance regions (which may be respectively expressed in terms of ``return periods" and ``return levels" in environmental sciences).
Given the (likely estimated) upper tail of a univariate distribution, one can begin with a small probability of interest and determine a threshold corresponding to the desired exceedance probability, or conversely begin with a very high quantile and determine the probability exceeding this quantile.
In the bivariate case, this one-to-one relationship no longer exists.
Given a risk region; that is, a region defined in terms of a specific bivariate extreme event occurring, EV methods have been devised to estimate the probability of such an event \citep[e.g.,][]{deHaan1998}.  
However in bivariate space, 
%the notion of exceedance region is ambiguous, and 
an exceedance region is not uniquely specified for a given probability.

%%return levels and return periods
%\note{Not sure if we need to talk about return levels and periods here or not.}
%Perhaps because estimated probabilities associated with rare events are so small that they are difficult to fathom, when assessing risk associated with natural disasters, people often speak of return levels and return periods.
%Assuming stationarity, an $r$-year return level (e.g., the so-called 100-year flood) is the level whose estimated probability of exceedance implies an expected time between events of $r$ (e.g., 100) years.
%Conversely, given a level of interest (often the level of an observed extreme event), people may give the return period (the expected time between such events) rather than the estimated probability of the event.
%In the univariate setting, a return level is simply a high quantile associated with the desired exceedance probability, and conversely, a return period is the exceedance probability associated with a quantile of interest.
%
%In the multivariate setting, return levels and return periods do not have a one-to-one relationship.
%Given a extreme region of interest, methods have been devised to estimate the probability (or equivalently the return period) of an event occurring in that region \citep[e.g.,][]{cooley2017} and others (old deHaan?).
%But a return level is harder as a probability does not uniquely define a region in bivariate space.

%equidensity curves
A familiar way to visually describe bivariate data is to draw contour lines corresponding to equal values of an estimated density function.
Typical methods will yield equidensity contours which form closed regions in $\mathbb{R}^2$.
However, equidensity contours may not be ideal for describing exceedance probabilities.
First, the contours' values correspond to density values rather than exceedance probabilities of the contour. % probability of `exceeding' the region formed by the contour.
Calculating associated exceedance probabilities would require integration of the density function over an oddly-shaped region.
More importantly, in most EV applications there is a direction of interest which is associated with impactful events.
We will assume that our data are oriented such that we are concerned when the variates take on their greatest values.
Notice that the data in Figure 1 reflect this orientation, and in particular the fire risk application has a ``dryness" variable which was constructed by negating humidity measurements.
An equidensity contour has no directional orientation.
%If an equidensity curve yields a closed region, only the portion of the region outside the curve in the direction of interest would be associated with actual risk.

%survival function based curves
Rather than equidensity contours, our tool will produce isolines such that the estimated survival probability of any point on the isoline is equal.
That is, if $\bm X=(X_1,X_2)^T$ takes values in $\mathbb{R}^2$ and $\hat {\bar F}_{\bm X}(\bm x) = \hat P(\bm X > \bm x) = \hat P(X_1 > x_1, X_2>x_2)$, $\bm x=(x_1,x_2)\in\mathbb{R}^2$, is its estimated survival function, then our isoline $\hat \ell_{\bm X}(p) := \{ \bm x \in \mathbb{R}^2 : \hat {\bar F}_{\bm X}(\bm x) = p \}$ for some exceedance probability of interest $p$.
By defining the isoline in terms of the survival function, we orient the exceedance region in the direction of interest, and tie the line directly to the specific notion of ``exceedance" given by the survival function.
Others have used isolines associated with probabilities of bivariate distributions.
\cite{salvadori2004} and \cite{marcon2017} draw isolines of extreme regions defined in terms of the survival function and additionally in terms of the cumulative distribution function.
The function {\tt qcbvnonpar} in the {\tt evd} package \citep{stephenson2002} in R draws isolines associated with the bivariate cumulative distribution.

%what's new
Our work relies on a dependence framework familiar to extremes, and is novel in that it is largely nonparametric.
Specifically, it begins with a nonparametric estimate of an isoline at a very high level, and then uses EV results to project to more extreme levels.
In contrast, \cite{salvadori2004} use parametric copula models and both \cite{marcon2017} and the {\tt qcbvnonpar} function \citep{stephenson2002} employ a semiparametric descriptor of bivariate extremal dependence.
Also importantly, we adapt our approach to draw isolines when data are determined to be asymptotically independent.
Asymptotic independence, described more completely in Section \ref{sec: background}, is a degenerate case for standard EV dependence frameworks, but data are frequently determined to exhibit asymptotic independence.
To our knowledge, all previous EV-based work to draw lines characterizing bivariate extreme behavior has assumed asymptotic dependence.
This includes the previously-cited work, and additionally \cite{cai2011, einmahl2009, coles94}.

%%%%%%%%%%%%%%%%%%%%%%%%%%%%%%%%%%%%%%%%%%%%%%%%%%
%MATH BACKGROUND	
%%%%%%%%%%%%%%%%%%%%%%%%%%%%%%%%%%%%%%%%%%%%%%%%%%
\section{Mathematical background for approach}
\label{sec: background}

%Outline
%1.  Ext lead in
%	Goal is model dependence in the tail -> rely on EVT
%	EV methods use small subset in tail.  
%	Assume asymptotic model
%	One approach is to use block maxima -> asy justtified approach is MVEVDs
% 	Because we want to model the fringe of daily dist, from a modeling approach seems unnatural to not use block max.
As we wish to draw contour lines at the utmost extent of the data and beyond, we must characterize dependence for the distribution's upper tail.
EV methods typically analyze only a small extreme subset of the available data in order that tail inference is not contaminated by non-extreme behavior.
Methods assume these largest values are well approximated by an asymptotically-justified model.
One approach is to use a subset of componentwise block (e.g., annual) maxima for which the limiting distributions are the class of multivariate extreme value distributions (MVEVDs).
Because we wish to visualize contours of the original data such as the daily data pictured in Figure \ref{fig: motivation}, rather than obtaining componentwise block maxima, we will use the largest values of the original data.

%2.  RV Reg var lead in
%	Informally, reg var heavy tailed.
%	Only described via tail.
%	Use is common in ext (finance) b/c of ties to MVEVDs laid out in Resnick and elsewhere.
Our method relies on the framework of regular variation to characterize the dependence in the tail of the distribution.
Informally, a bivariate regularly varying random vector is one whose joint distribution has a heavy tail, implying that the tail decays like a power function.
Because the definition, given below, only describes behavior in the joint tail, and because only extreme data are used for inference, the data from the distribution's bulk does not influence inference.
More importantly, the fundamental dependence structure of multivaritate regular variation can be directly linked to that of the MVEVDs \citep[Section 5.4.2]{resnick1987}, justifying its use for extremes.

%3.  Reg var definition
%	Extrapolation motivation.
%	Here, relation is in terms of EVI $\xi$ rather than index of reg variation $\alpha = 1/\xi$ as readers may be more familiar with $\xi$ from other environmental apps of extremes.
Formally, a nonnegative bivariate random vector $\bm Z$ is regularly varying if there exists a renormalizing sequence $b_n \rightarrow \infty$ as $n \rightarrow \infty$, and a measure $\nu$ on the space ${\cal C} = [0, \infty]^2 \setminus \bf 0$, such that as $n \rightarrow \infty$
\begin{equation}
  \label{eq: regVarDefn}
  n P \left( \frac{\bm Z}{b_n} \in A \right) \rightarrow \nu(A),
\end{equation}
for any $\nu$-continuity set $A \subset {\cal C}$.
The normalizing sequence $b_n$ is regularly varying with extreme value index $\xi > 0$; that is $b_n = n^{\xi} L(n)$ where $L(n)$ is a slowly varying function \citep{resnick2007}.
The limiting measure $\nu$ has the property such that
\begin{equation}
  \label{eq: scaling}
  \nu(sA) = s^{-1/\xi}\nu(A),
\end{equation}
for any scalar $s > 0$ and $A \subset {\cal C}$.
%From (\ref{eq: scaling}), we see the power-law behavior of the tail, and $\xi$ is termed the extreme value index.
We parametrize in terms of $\xi$, rather than the index of regular variation $\alpha = 1/\xi$, as readers may be more familiar with this parameter from other environmental extremes work.
Larger values of $\xi$ indicate heavier tails, and (\ref{eq: scaling}) is useful for extrapolating further into the tail.

%4.  Asy indep/dep
%	In practice, unnatural
%	MVEVDs -> indep
%	Reg var -> prob of set in interior of cone is zero.
%	Our case is really hard!
Asymptotic (in)dependence is a notion that describes fundamental bivariate tail behavior.
Let $\bm X = (X_1, X_2)^T$ be a bivariate vector (not necessarily regularly varying) with univariate marginal cumulative distribution functions $F_{X_1}$ and $F_{X_2}$.
Define 
$$
  \chi = \lim_{u \rightarrow 1} P(F_{X_1}(X_1) > u \mid F_{X_2}(X_2) > u).
$$
$\bm X$ is deemed asymptotically independent if $\chi = 0$, and is deemed asymptotically dependent otherwise.
Intuitively, asymptotic dependence implies that the two variates can obtain their largest values simultaneously.

A model will either be asymptotically dependent or independent, and it is essential for estimating joint tail probabilities that the selected model correctly captures the behavior exhibited by the data.
Regular variation is a useful modeling framework for describing tail dependence under asymptotic dependence.
Many familiar multivariate models are asymptotically independent, including the Gaussian and most copula models, and these will underestimate joint exceedance probabilities estimated by extrapolation into the tail if applied to data which are asymptotically dependent.
However, asymptotic independence is a degenerate case for regular variation (as well as for the MVEVDs).
If $\bm Z$ is regularly varying and asymptotically independent, then for any set $A \subset {\cal C}$ which does not include a portion of the axes, $\nu(A) = 0$.

%5.  Hidden regular variation. 
%	beyond scope of work
%	intuition, normalizing sequence is too fast
%	replace with something else
%	motivation for survival function-based curves
Ledford and Tawn (\citeyear{ledford1996, ledford1997}) were among the first to extend the regular variation framework to account for tail dependence in the asymptotically independent setting, and \cite{resnick2002} further formalized ideas via the concept of hidden regular variation.
%A full treatment of hidden regular variation is beyond the scope of this work, but 
An intuitive explanation of asymptotic independence is that for sets $A \subset {\cal C}$ which do not include points on the axes, the renormalizing sequence $\{ b_n \}$ in (\ref{eq: regVarDefn}) grows too rapidly, and the resulting limit is 0.
However, hidden regular variation obtains nontrivial convergence for such sets by using a lighter-tailed normalizing sequence $\{ b_n^0 \}$ with coefficient of tail dependence $\eta < \xi$:
\begin{equation}
  \label{eq: convHRV}
  nP \left( \frac{\bm Z}{b_n^0} \in A \right) \rightarrow \nu_0(A).
\end{equation}
The scaling property for sets $A$ bounded away from the axes and scalar $s > 0$ is
\begin{equation}
  \label{eq: scalingHRV}
  \nu_0(sA) = s^{-1/\eta} \nu_0(A),
\end{equation}
whereas the scaling property (\ref{eq: scaling}) continues to hold for sets including a portion of the axes.
A model property of hidden regular variation is a abrupt transition between (\ref{eq: regVarDefn}) and (\ref{eq: convHRV}) for sets which include portions of the axes and sets which do not \citep[c.f.,][]{das2014, weller2013}.
%This awkward transition is illustrated in the context of simulation in \cite{das2014}.
%who state, ``Both multivariate regular variation and hidden regular variation are asymptotic models with curious properties which are often ignored or misinterpreted when attempting to generate finite samples exhibiting such properties."

%6.  Copula-like things
%	Reg var assumes a heavy-tailed marginal and our data don't display it.
%	Like copula approaches (cite Nelson?), our approach separates dependence modeling from marginal behavior.
%	Diff in that instead of [0,1], we assume (after trans) heavy-tailed with index 1 (cite Ledford and Tawn?)
%	Get to $\eta$.
The regular variation framework described above requires that each univariate marginal distribution be heavy-tailed with extreme value index $\xi$.
However, when viewed as a copula, the dependence framework can be used to model data which are not heavy-tailed.
Like copula modeling approaches and much extremal dependence modeling work, our approach assumes a dependence framework after transformation to a convenient marginal.
As \cite{ledford1996}, we choose to transform so that each marginal can be assumed to be regularly varying with extreme value index $\xi = 1$.
Marginal transformation can be defended by Proposition~5.10 of \cite{resnick1987} which states that the domain of attraction of a MVEVD is preserved under monotonic marginal transformation, and this result can be interpreted as the extremes equivalent of Sklar's theorem from copula theory \citep{nelsen2006}.

%%%%%%%%%%%%%%%%%%%%%%%%%%%%%%%%%%%%%%%%%%%%%%%%%%
%ASYMPTOTIC DEPENDENT CASE
%%%%%%%%%%%%%%%%%%%%%%%%%%%%%%%%%%%%%%%%%%%%%%%%%%
\section{Asymptotic dependent case:  procedure and Santa Ana example}
\label{sec: asyDepCase}

%Santa Ana data
We use the data for the Santa Ana weather regime to illustrate the isoline procedure in the case of asymptotic dependence.
Hourly data were obtained from the HadISD dataset \citep{dunn2012} for the March AFB station\footnote{Station number 722860-23119.}, which lies in Riverside County, CA and whose data record shows a correspondence with known Santa Ana events such as the Cedar and Witch fires.
The data span the years 1973-2015 and we restrict our attention to the months of September, October, and November as these are months for which Santa Ana-driven fires are most prevalent.
Details and motivation for construction of the daily windspeed and dryness time series from this hourly data are given in \cite{cooley2017}.
The dataset contains $n=3902$ observations.
Let $\bm X_t = (X_{t,1}, X_{t,2})^T$ denote the random vector representing windspeed and dryness on day $t$ and let $\bm x_t$ denote the corresponding observation. %, and $\bm X_t$ is assumed to be identically distributed with $F_{\bm X}$ for all $t$.
Asymptotic dependence is a reasonable starting assumption for the Santa Ana data since the weather regime leads to conditions which are both very dry and windy. 
The {\tt chiplot} function of  {\tt R}'s {\tt evd} library (not shown) indicates that $\hat \chi \approx 0.3$, implying that conditions are at their driest levels about 30\% of the time when windspeeds are at their highest levels.
%Because the risk of fire is associated with dry conditions, the dryness variable was constructed from negated humidity measurements giving us the correct directional orientation.

%%chiplot
%To further confirm the assumption of asymptotic dependence, the left panel of Figure \ref{fig: santaAnaFigs} shows an empirical estimate for $\chi$.
%Plots of $\hat \chi$ are rarely absolutely conclusive for asymptotic dependence, but this plot does not show the characteristic drift of $\hat \chi$ toward zero for increasing values of $u$ typically seen with asymptotically independent variables.
%As $u \rightarrow 1$, values for $\hat \chi$ seem to be slightly increasing to a level somewhat above 0.3 before the sample size becomes so small that estimates become very uncertain.
%Interpreting $\hat \chi$, conditions are at their driest levels about 30\% of the time when windspeeds are at their highest.

%drawing the base contour line
The first step in the procedure is to nonparametrically construct a ``base" isoline.
Let $p_{base}$ be the exceedance probability selected for this base isoline;
$p_{base}$ should be small enough such that extreme dependence is well represented, and yet should be large enough for the nonparametric procedure to have adequate data.   
Our method for constructing the base isoline begins with a Gaussian-kernel based estimate of the cumulative distribution function similar to that proposed by \cite{liu2008}. 
Kernel bandwidth can either be specified by the user, or one can employ an automated bandwidth selection tool; we use the {\tt bandwidth.nrd} tool employed by the {\tt kde2d} density estimation tool in {\tt R}'s {\tt MASS} library. 
The survival function's value is estimated on a fine grid spanning the range of the data, and $\hat {\bar F}_{\bm X}$ is monotonically decreasing by construction.
The base contour line $\hat \ell_{\bm X}(p_{base}) = \{ \bm x \in \mathbb{R}^2 : \hat F_{\bm X}(\bm x) = p_{base} \}$ can be drawn via standard interpolation methods.
For the Santa Ana data we let $p_{base} = 0.01$, and the left panel of Figure \ref{fig: santaAnaFigs} shows the base isoline in red.

\begin{figure}[ht]
  \begin{center}
    \includegraphics[scale = .75]{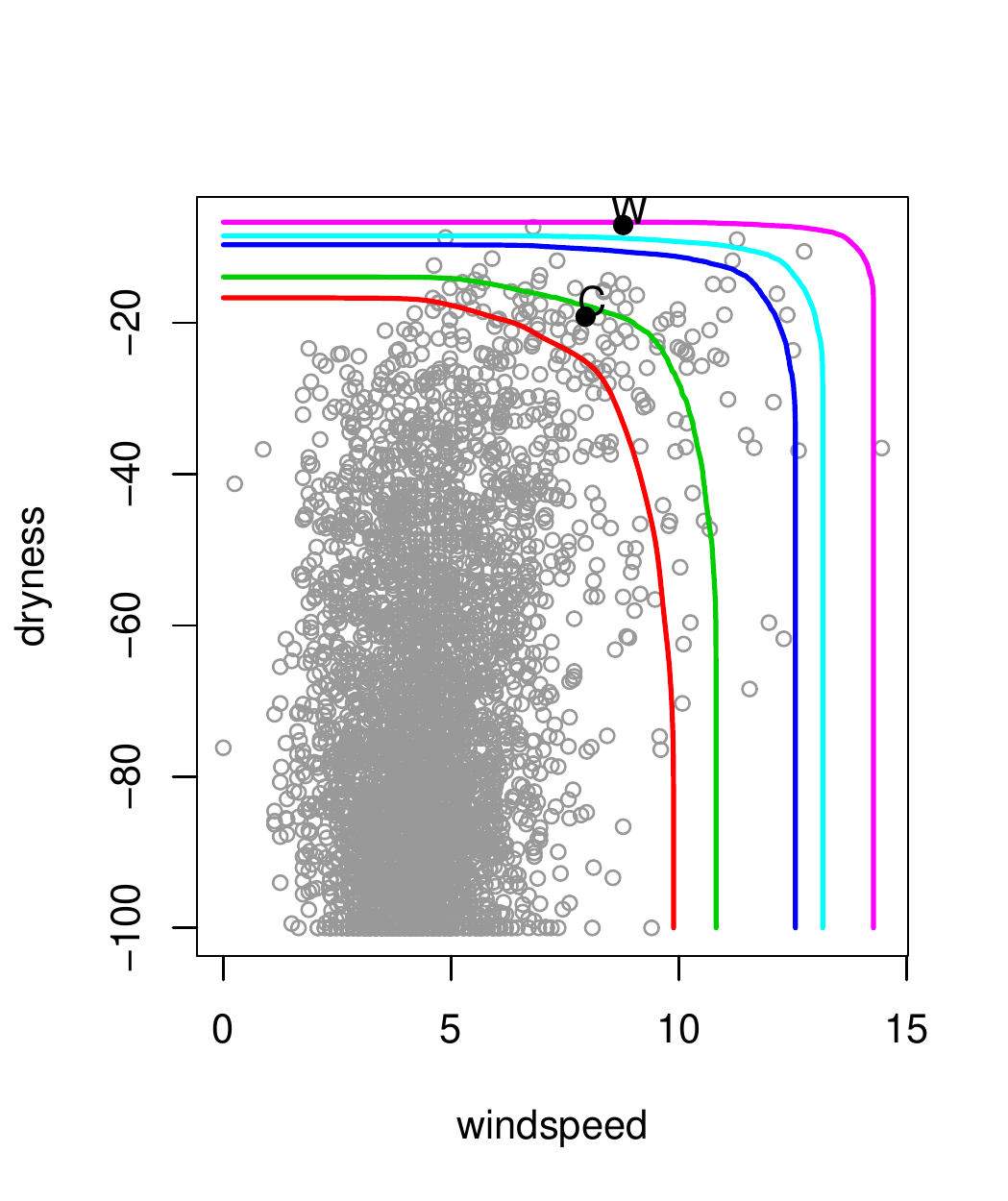}
    \includegraphics[scale = .75]{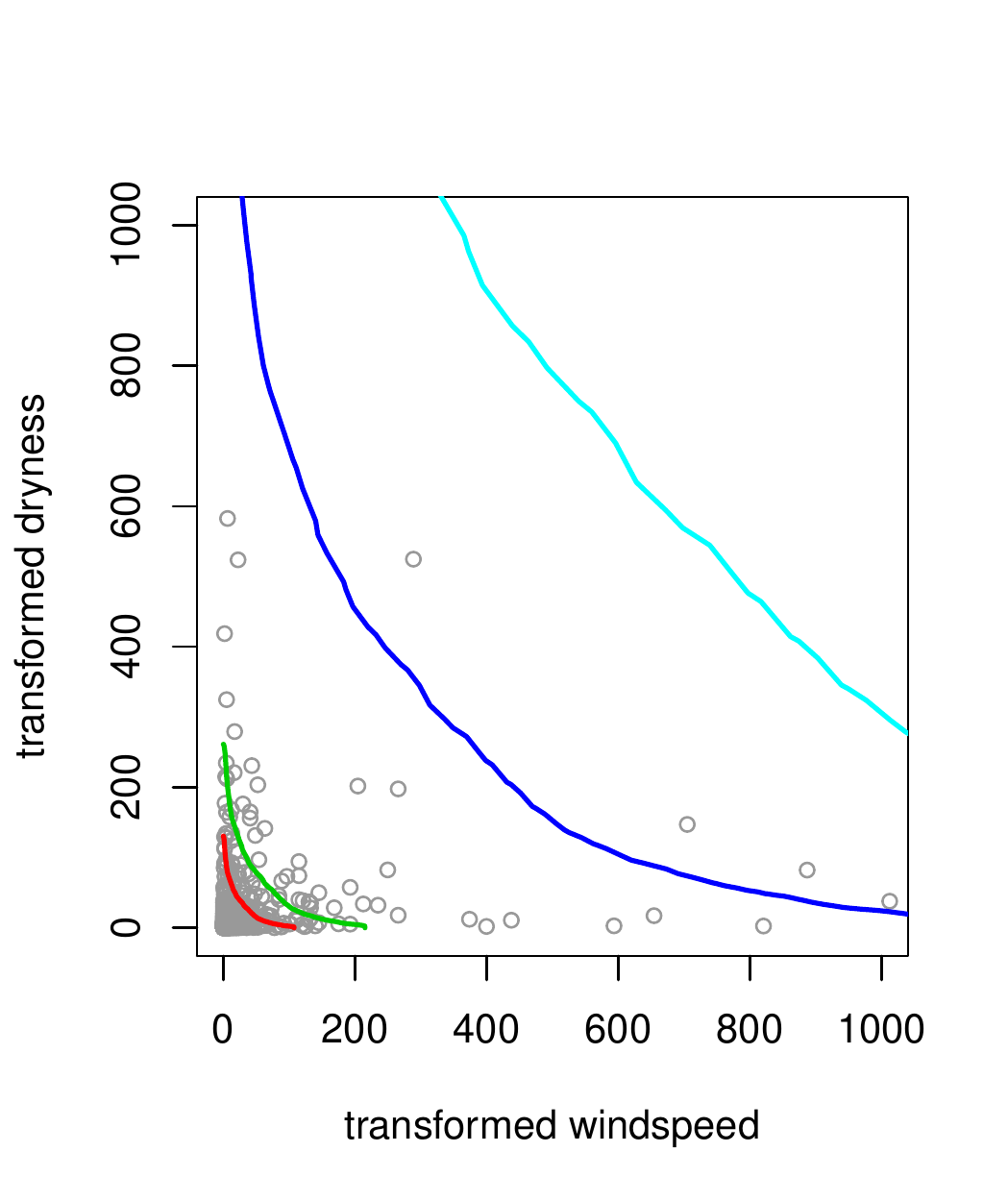}
  \end{center}
  \caption{Plot of the Santa Ana data and estimated survival function isolines on the original scale (left), and on the transformed scale (right).  Estimated isolines correspond to survival probabilities of 0.01, 0.005, 0.001, 0.0005, and 0.0001.}
  \label{fig:  santaAnaFigs}
\end{figure}

%transforming marginals
We intend to use (\ref{eq: scaling}) to extrapolate the base contour line to levels corresponding to smaller survival probabilities.
%However, unless the data are assumed to be regularly varying, transformation is required.
However, marginal transformation is required as the data itself is not regularly varying; both the variables in the Santa Ana data set are found to have bounded upper tails. 
%, and the regular variation framework outlined in Section \ref{sec: background} requires the variables to be heavy-tailed, 
%Much like copula approaches, we assume that the fundamental dependence structure is unchanged by monotonic marginal transformation, and \citep[Prop.  XX]{resnick87} showed that the domain of attraction of a multivariate extreme value distribution is \note{fill in the right words here.}
Let $F_k$ denote the marginal distribution for $X_{t,k}$, $k = 1,2$.
Our marginal transformation procedure begins by constructing a linearly interpolated empirical cumulative distribution function $\hat F^{emp}_k(x)$, over the range of the data of each marginal.
To allow extrapolation further into the tail, we additionally fit a generalized Pareto distribution $\hat F^{gpd}_k(x)$ above a high threshold $x_{thold, k}$ for each marginal.
Since our contour lines will take on both large and small values of each variate, we construct a smooth transition between the two marginal estimates.
Define a weight function $w_k(x)$ where $w_k(x) = 0$ for $x \leq x_{thold, k}$, $w_k(x)$ is monotonically increasing from 0 to 1 in the range $x_{thold, k} < x < x_{thold+,k}$ and $w_k(x) = 1$ for $x > x_{thold+,k}$.
Letting $\hat F_{k}(x) = (1 - w(x)) \hat F^{emp}_k(x) + w(x)  \hat F^{gpd}_k(x)$, we then construct a marginal transformation function
$$
  T_k(x) = -\log^{-1}(\hat F_k(x)).
$$
Thereby $\bm Z_t = T(\bm X_t) = (T_1(X_{t,1}), T_2(X_{t,2}))^T$ can be assumed to be regularly varying with $\xi = 1$.
We employ the function $w_k(x) = (\sin(\pi(x - x_{k,thold})/(x_{k, thold+} - x_{k,thold}) - \pi/2)+1)/2$.
$\hat F_k(x)$ must be verified to be monotonically increasing after smoothing.
After checking threshold diagnostics, we set $x_{thold,k} = q_{0.97}$ (the 0.97 quantile) and $x_{thold+,k} = q_{0.98}$ for both marginal distributions of the Santa Ana data.
%Letting $\bm x_t$ represent the observation from day $t$, the right plot of Figure \ref{fig: santaAnaFigs} shows the transformed data $T(\bm x_t), t = 1, \ldots, n$.

%projecting in Frechet scale
%Let $\{ \ell_{p_{base}}^{RV(\alpha)} = \{ \bm z \in R^2 : \bm z = T(\ell_{p_{base}}) \}$.
%To project the base isoline, begin by assuming that the sample size $n$ is fixed and large enough such that for any set $A \subset {\cal C}$, (\ref{eq: scaling}) holds approximately.
%\begin{eqnarray*}
%  n P \left( \frac{\bm Z_t}{n^{1/\alpha} L(n)} \in A \right) &\approx& \nu(A)  \\
%  \Rightarrow P( \bm Z_t \in A_*) &\approx& k^\alpha \nu(A_*), 
%\end{eqnarray*}
%where $A_* = n^{1/\alpha} L(n) A$, and and $k = L(n)$ for the fixed $n$.
%$A_*$ must consist of large values, since $A$ is bounded away from $\bm 0$ and $n$ is large implies $\| \bm z \|$ must be large for all $\bm z \in A_*$.
%Thus, for large sets $A_*$ and for $s > 1$, (\ref{eq: scaling}) implies
%\begin{equation}
%  \label{eq: probScaling}
%  P(\bm Z \in sA_*) \approx s^{-\alpha} P(\bm Z \in A_*).
%\end{equation}
Let $\hat \ell_{\bm Z}(p_{base}) =T(\hat \ell_{\bm X}(p_{base}))$.
%The right plot of Figure \ref{fig: santaAnaFigs} shows the transformed data and base isoline in red.
To project this transformed base isoline to higher levels corresponding to smaller exceedance probabilities, begin by assuming that the sample size $n$ is fixed and large enough such that for any set $A \subset {\cal C}$, (\ref{eq: regVarDefn}) holds approximately:
\begin{eqnarray*}
  n P \left( \frac{\bm Z_t}{n L(n)} \in A \right) &\approx& \nu(A)  \\
  \Rightarrow P( \bm Z_t \in A_*) &\approx& k \nu(A_*), 
\end{eqnarray*}
where $A_* = n L(n) A$, and $k = L(n)$ for the fixed $n$.
$A_*$ is understood to consist of large values, since $A$ being bounded away from $\bm 0$ and $n$ being large implies $\| \bm z \|$ must be large for all $\bm z \in A_*$.
Thus, for large sets $A_*$ and for $s > 1$, (\ref{eq: scaling}) implies
\begin{equation}
  \label{eq: probScaling}
  P(\bm Z \in sA_*) \approx s^{-1} P(\bm Z \in A_*).
\end{equation}
By construction, $P(\bm Z_t \in [\bm z, \bm \infty)) = p_{base}$ for any $\bm z \in \hat \ell_{\bm Z}(p_{base})$.  
From (\ref{eq: probScaling}), setting $s = p_{base}/p_{proj}$ for any $p_{proj} < p_{base}$, then $P(\bm Z_t \in [s\bm z, \bm \infty)) = p_{proj}$ for any $\bm z \in \hat \ell_{\bm Z}(p_{base})$.
Thus on the transformed scale, we can construct $\hat \ell_{\bm Z}(p_{proj}) = s \hat \ell_{\bm Z}(p_{base})$.

The right panel of Figure \ref{fig: santaAnaFigs} shows the data after transformation: $\bm z_t = T(\bm x_t), t = 1, \ldots, n$; the transformed base isoline in red, and the projected isolines corresponding to $p_{proj} = 0.005, 0.001, 0.0005$.
Scatterplots on the transformed scale can be difficult to interpret due to unfamiliarity with the extremely heavy tail when $\xi = 1$ \citep{cooley2017}.
%Plotting on the transformed scale can be awkward due to the extremely heavy tail when $\alpha = 1$ \citep{cooley2017}.
Our plot window was restricted to $[0,1000]^2$ which does not contain all of the data after transformation, nor does it contain any of the isoline $\hat \ell_{\bm Z}(0.0001)$ which nevertheless was produced.
To produce isolines on the original scale, simply reverse the transformation:  $\hat \ell_{\bm X}(p_{proj}) = T^{-1}( \hat \ell_{\bm Z}(p_{proj}) )$.
The left panel of Figure \ref{fig: santaAnaFigs} shows these isolines.
We see that the combination of windspeed and dryness seen at this weather station on the day corresponding to the Cedar Fire was not extremely rare, as this point is below $\hat \ell_{\bm X}(0.005)$.
The conditions at this station on the day corresponding to the Witch Fire were quite rare as this point lies between $\hat \ell_{\bm X}(0.0005)$ and $\hat \ell_{\bm X}(0.0001)$.

%%%%%%%%%%%%%%%%%%%%%%%%%%%%%%%%%%%%%%%%%%%%%%%%%%
%ASYMPTOTIC INDEPENDENT CASE
%%%%%%%%%%%%%%%%%%%%%%%%%%%%%%%%%%%%%%%%%%%%%%%%%%
\section{Asymptotic Independence case: procedure and Karachi example}
\label{sec: asyIndepCase}

%data
%	background, where from, years
%	Exceptionally hard case--negative dependence.
We use the Karachi heat wave data to illustrate the method in the asymptotic independence setting.
The data are again a HadISD dataset \citep{dunn2012}, this time for the Karachi Airport station.\footnote{Station number 417800-99999.}
The original data span 1973-2015 and are nominally hourly, although the recording interval was every four hours until 1992.
For each day, we retain the temperature (converted to Fahrenheit) and relative humidity both at the time of the maximum heat index value which itself is a function of temperature and humidity.
We examine data from the months from April to October, as these were the months during which a heat index value exceeded the 0.95 empirical quantile.
Time series plots of temperature and humidity still display seasonality, as temperatures tend to increase in April and May, peak in June, dip in the monsoon months of July and August, then increase again in September before decreasing in October.
Relative humidity tends to be higher in the monsoon months.
This seasonality makes interpretation of the isolines more difficult, as one must think of the isolines representing exceedance probabilities of the distribution of these two variables integrated over the warm months.
The data set contains $n = 8963$ data points.  

The Karachi data are clearly asymptotically independent.
In fact these data exhibit negative association: days with highest temperature are days with lower humidity, and conversely, the days with highest humidity are days with lower temperatures.
This negative association is to be expected as moisture in the air mitigates temperature.
Despite this negative relationship, there is a need to characterize the relationship in the tail of these two variables as human health is adversely affected when both temperature and humidity are high.

The procedure in the asymptotic independent setting begins in the same manner as before.
Let $\bm X_t = (X_{t,1}, X_{t,2})$ represent the temperature and relative humidity from day $t$.
First, a base isoline $\hat \ell_{\bm X}(p_{base})$ is nonparametrically estimated using the original data in the same manner as before.
The left panel of Figure \ref{fig: karachiFigs} shows this base isoline in red.
As before, the data require transformation before regular variation can be assumed.
We obtain $\hat F_k(x)$ for $k = 1,2$, from it define the marginal transformations $T_k$, and again let $\bm Z_t = T(\bm X_t)$ which we assume to be regularly varying with index $\xi = 1$ on ${\cal C}$.
The transformed data $\bm z_t = T(\bm x_t)$ and transformed base isoline $\hat \ell_{\bm Z}(p_{base}) = T(\hat \ell_{\bm X}(p_{base}))$ are shown in the right panel of Figure~\ref{fig: karachiFigs}.
Note that this panel's range is set to $[0,50]^2$ in order to show the behavior of points in the interior of ${\cal C}$ and there are many large points near the axes which are beyond this range.

\begin{figure}[ht]
  \begin{center}
    \includegraphics[scale = .75]{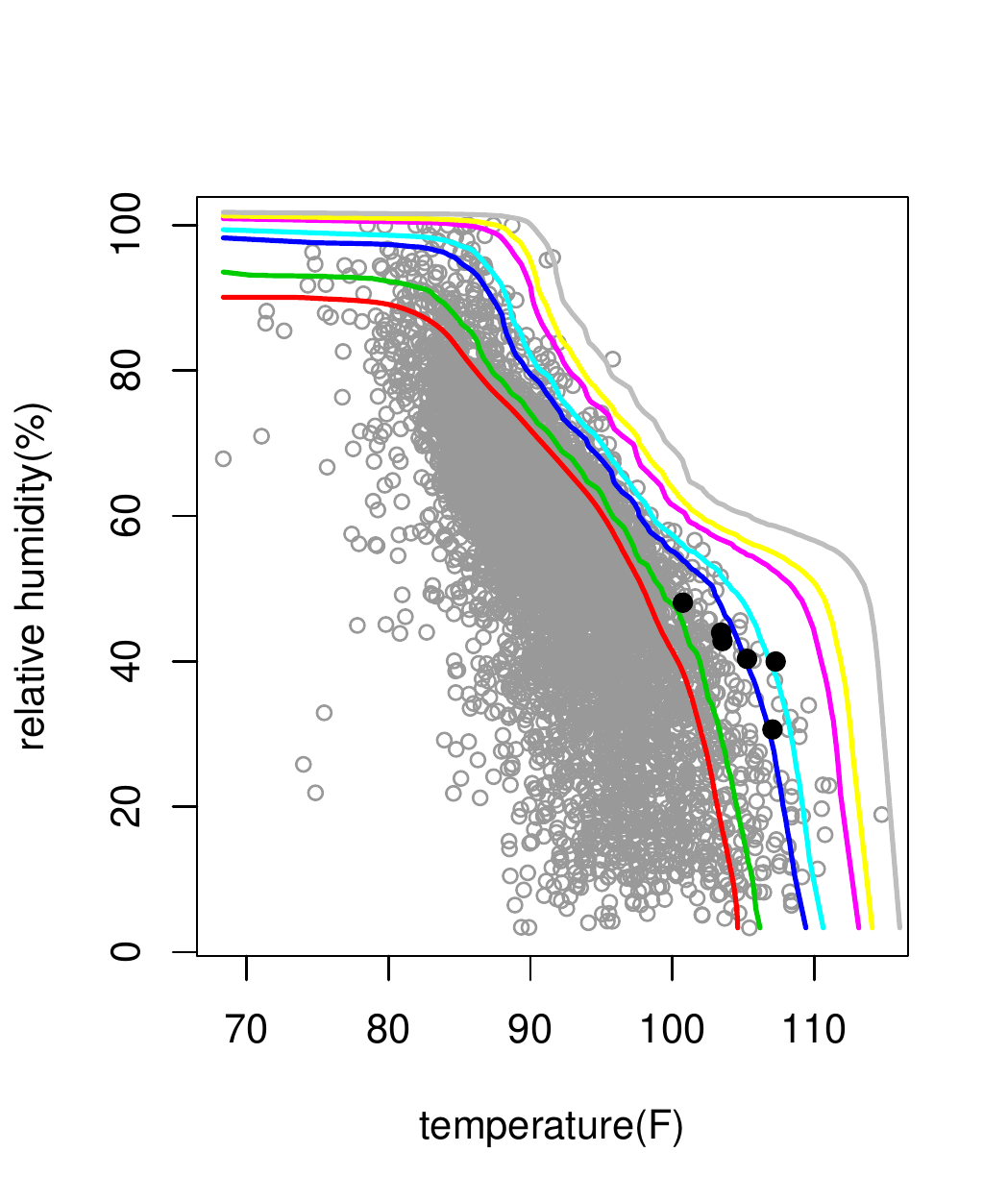}
    \includegraphics[scale = .75]{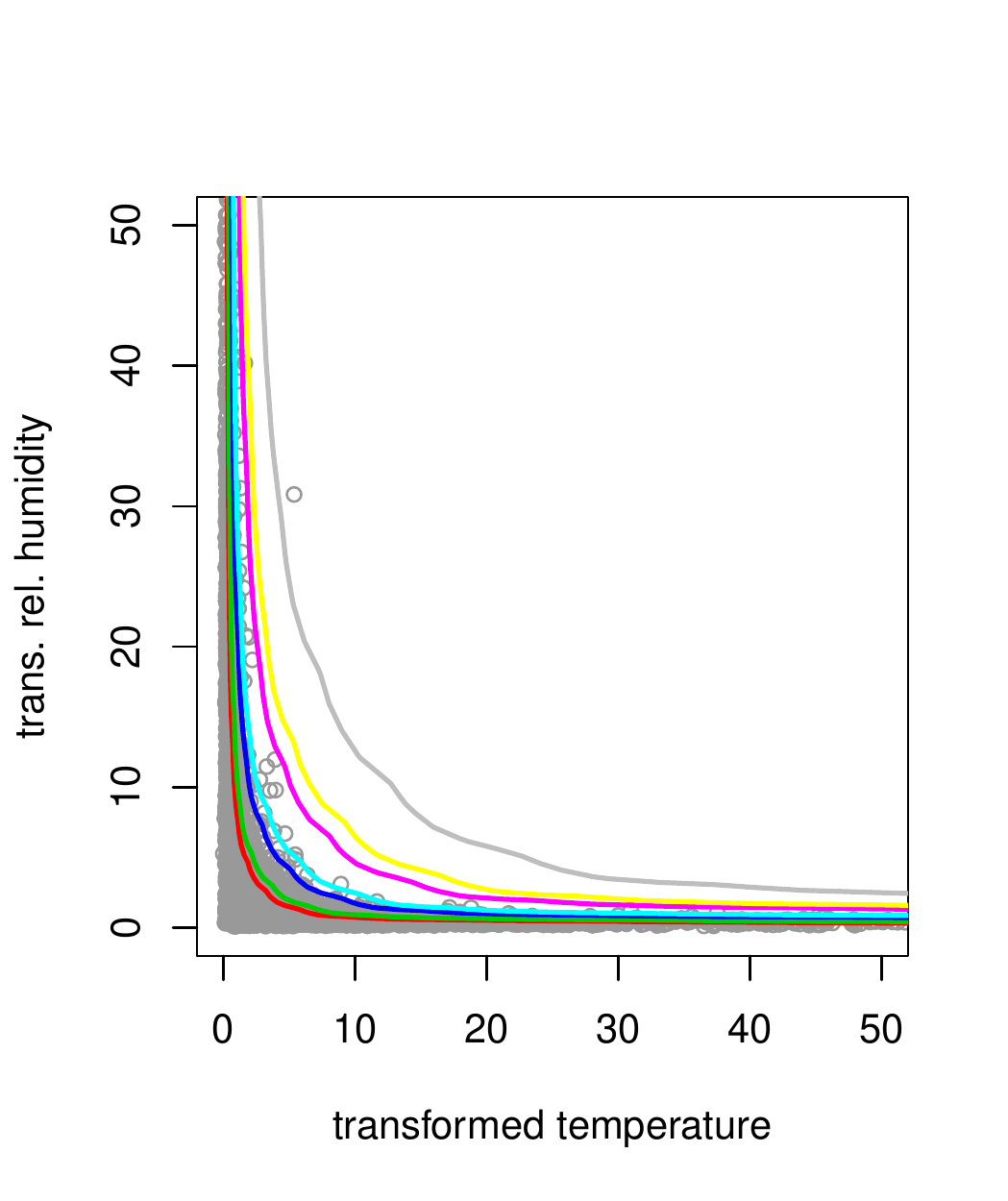}
  \end{center}
  \caption{Plot of the Karachi data and estimated survival function isolines on the original scale (left), and on the transformed scale (right).  Estimated isolines correspond to survival probabilities of 0.01, 0.005, 0.001, 0.0005, 0.0001, 0.00005, and 0.00001.}
  \label{fig:  karachiFigs}
\end{figure}

%estimating eta
%note that dependence implies 1/2
Assuming $\bm Z_t$ exhibits hidden regular variation on the interior of ${\cal C}$, then  note $Z_t^{(min)} = \min(Z_{t,1}, Z_{t,2})$ should be regularly varying with index $\eta$ \citep{ledford1996}.
We estimate $\eta$ via a Hill estimator \citep{hill1975} applied above a threshold corresponding to the 0.98 empirical quantile of $z_t^{(min)}$ of the Karachi data, and we obtain $\hat \eta = 0.20$.
This value indicates negative correspondence in the tail as independence between the variates would implies $\eta = 1/2$ when $\xi = 1$.
A Hill plot was produced which indicated that this quantile was reasonable for estimating $\eta$.

%procedure up to projection
%	segue with need to project with eta
%	choice of survival function set
We next wish to use the scaling properties of regular variation and hidden regular variation to produce isolines corresponding to smaller exceedance probabilities.
Because we have chosen our isolines to be based on joint survival regions, our procedure is able to utilize hidden regular variation to scale the interior.
However, the interface between the first-order regular variation and hidden regular variation that occurs at the axes presents a modeling challenge.
If $\bm z \in \hat \ell_{\bm Z}(p_{base})$ is in the interior of ${\cal C}$, the asymptotic theory says that (\ref{eq: scalingHRV}) is the correct relationship to use; however, if $\bm z$ lies on an axis, then (\ref{eq: scaling}) should be used.
This abrupt change would result in discontinuous isolines at the axes.

Because the abrupt transition between regimes described by the asymptotic theory is not reflected in the data, we propose smoothing the transition between scalings (\ref{eq: scaling}) and (\ref{eq: scalingHRV}).
Let $\bm z^{(base)} = (z^{(base)}_1, z^{(base)}_2) \in \hat \ell_{\bm Z}(p_{base})$.
For some smoothing parameter $\beta$, let $m_i = 1 - \big( z^{(base)}_i/(z^{(base)}_1 + z^{(base)}_2)  \big)^\beta$, and let $\eta_i(\bm z^{(base)}) = m_i \hat \eta + (1 - m_i)$ for $i = 1,2$.
To construct an isoline to correspond with an exceedance probability $p_{proj} < p_{base}$, let $s = p_{base}/p_{proj}$ as before.
Consider $\bm z^{(proj)} = (s^{\eta_1(\bm z^{(base)})} z^{(base)}_1, s^{\eta_2(\bm z^{(base)})} z^{(base)}_2)$.
If $\bm z^{(base)}$ is sufficiently away from the axes such that $m_i \approx 1$ for $i = 1,2$, then
$$
  P(\bm Z_t > \bm z^{(proj)}) \approx P(\bm Z_t > s^{\hat \eta} \bm z^{(proj)}) \approx (s^{\hat \eta})^{-1/\eta} P(\bm Z_t > \bm z^{(base)}) \approx p_{proj},
$$
where the second approximation comes from (\ref{eq: scalingHRV}).
On the other hand consider the case where $\bm z^{(base)}$ lies on the axis.
Suppose $z^{(base)}_1 = 0$.
$$
  P(\bm Z_t > \bm z^{(proj)}) = P(Z_{t,1} > 0, Z_{t,2} > s z^{(base)}_2) \approx s^{-1} P(\bm Z_t > \bm z^{(base)}) = p_{proj},
$$
where the approximation follows from (\ref{eq: scaling}).
For $\bm z^{(base)}$ near the axes, these two cases are weighted with the weight depending on the value of $\beta$ selected by the investigator.
As $\beta \rightarrow \infty$, one approaches the discontinuous transition between regimes and the projection of the isoline is scaled primarily by the coefficient of tail dependence $\eta$ even near the axes.
As $\beta$ decreases, smoothing is increased, and the first-order regular variation influences behavior further from the axes.

If $\hat {\bar F}_{\bm X}(\bm x)$ is decreasing, isolines of survival probabilities have negative slopes.
We show in the appendix that projected isolines produced by the above smoothing procedure retain this property.

Using diagnostics to be explained in Section \ref{sec: diag}, we selected $\beta = 200$ and produced isolines $\hat \ell_{\bm Z}(p_{proj})$ for $p_{proj} = 0.005, 0.001, 0.0005, 0.0001, 0.00005, 0.00001$, which are shown in the right panel of Figure \ref{fig: karachiFigs}.
Inverting the marginal transformation yields the estimated isolines on the original scale shown in the left panel of Figure \ref{fig: karachiFigs}.
The conditions recorded at the Karachi airport on the dates between June 18 and June 23, 2015 were all rare.
The points corresponding to these dates all exceed $\hat \ell_{\bm X}(0.005)$, three exceed $\hat \ell_{\bm X}(0.001)$ and one exceeds $\hat \ell_{\bm X}(0.0005)$.
%Undoubtedly, the fact that these extreme days occurred in succession contributes to the experienced consequences to human health of this event.
%However, the plot may also give some indication that impacts do not necessarily align with rarity.  
%The isolines show several points with estimated survival probabilities less than those occurring during the June 2015 heat wave.  
%Comparing the survival probability contours with known impactful events may provide practitioners knowledge of what regions of the bivariate space have the potential for largest impact.

%%%%%%%%%%%%%%%%%%%%%%%%%%%%%%%%%%%%%%%%%%%%%%%%%%
%DIAGNOSTICS
%%%%%%%%%%%%%%%%%%%%%%%%%%%%%%%%%%%%%%%%%%%%%%%%%%
\section{A Diagnostic Plot and Bootstrap Uncertainty}
  \label{sec: diag}

%Many diagnostic plots commonly used for extremes can be applied to different portions of our procedure.
%Mean residual life plots \citep[][Section 4.3.1]{coles2001} and Hill plots \citep[][Section 4.2]{beirlant2004} can be used to help choose thresholds for fitting the marginal GPDs, and qq plots can be used to assess the data transformation.
%A Hill plot can also help assess the estimate for $\eta$.

We propose a plot to assess whether a projected isoline is sensible, and this plot can also help in selecting an appropriate smoothing parameter $\beta$.
Since we presume  $\hat {\bar F}_{\bm X}(\bm x) = p$ for any $\bm x \in \hat \ell_{\bm X}(p)$, we propose plotting the empirical survival probability for a number of points along a projected contour line.
To help quantify the expected uncertainty of these empirical probabilities, for any point $\bm x \in \hat \ell_{\bm X}(p)$, let $B(\bm x) = \sum_{t = 1}^n \mathbb{I}(\bm X_t \in [\bm x, \bm \infty))$.
Assuming $B(\bm x) \sim {\rm Binomial}(n, p)$, we find the smallest interval $(n_1, n_2)$, $n_i \in \mathbb{Z}$, such that $P(B(\bm x) \in [n_1, n_2]) \geq 0.95$.
We then report the interval $n^{-1}[n_1, n_2]$ which can be compared to the empirical probabilities $B(\bm x)/n$.
For two points $\bm x_1, \bm x_2 \in \ell_{\bm X}(p)$, $B(\bm x_1)$ and $B(\bm x_2)$ are not independent as the regions $[\bm x_1, \bm\infty]$ and $[\bm x_2, \bm \infty]$ will overlap, with extensive overlap if $\bm x_1$ and $\bm x_2$ are close to one another.
The aforementioned interval does not account for this dependence.

Figure \ref{fig: betaPlots} shows empirical survival probability plots for the Santa Ana data on the left, and plots for the Karachi data for two different values of $\beta$ center and right.  
The plotted empirical survival probabilities clearly show dependence.
The plot for the Santa Ana data is for $p = 0.001$, and all empirical exceedance probabilities fall within the uncertainty interval.
The Karachi plot in the center is for $\beta = 200$ and $p = 0.0005$.  
While some empirical probabilities fall both above and below the uncertainty interval, the values are close to the interval bounds.
This is not the case for the right plot, where $\beta = 1000$.  
The empirical probabilities on the far right and far left clearly exceed the upper bound of the uncertainty interval by a large amount indicating a clear discrepancy of this isoline.
With such a large value for $\beta$, the transition from regular variation in the interior to the first-order regular variation on the axes is too abrupt, and $\eta$ overly influences the scaling near the axes, resulting in too many observed exceedances.
Returning to the center plot, we tried changing both $\beta$ and reducing the exceedance probability of the base isoline, but were unable to qualitatively improve the performance beyond what is demonstrated in this plot.

\begin{figure}[ht]
  \begin{center}
    \includegraphics[scale = .7]{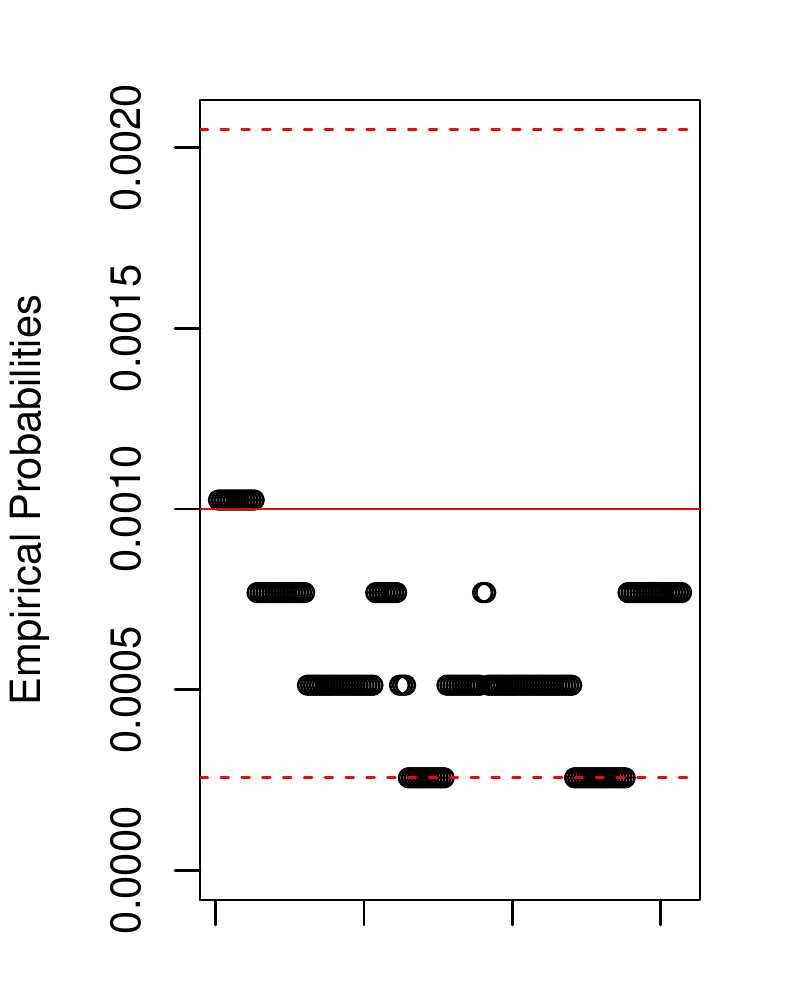} \hspace{-.4 in}
    \includegraphics[scale = .7]{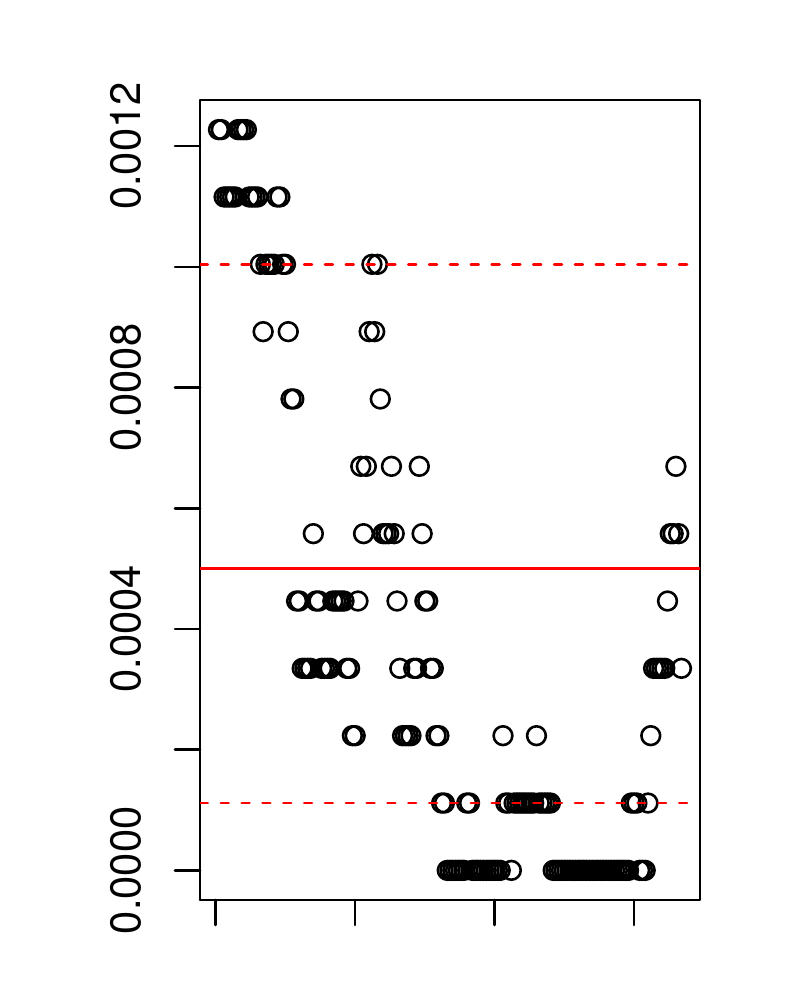} \hspace{-.4 in}
    \includegraphics[scale = .7]{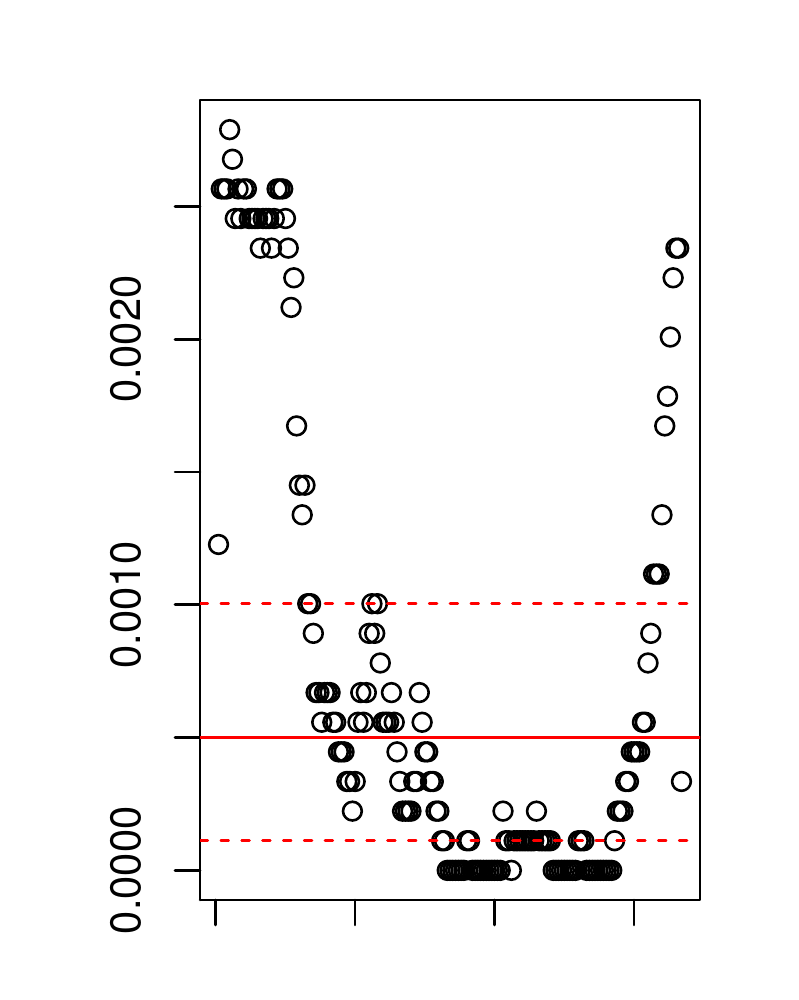}
  \end{center}
  \caption{Plot of empirical survival probabilities for points along the estimated isoline.  Left:  Santa Ana data, for $\hat\ell_{\bm X}(0.001)$.  Center:  Karachi data for $\hat \ell_{\bm X}(0.0005)$, with $\beta = 200$.  Right:  Karachi data for $\hat \ell_{\bm X}(0.0005)$, with $\beta = 1000$.  Solid line corresponds to target isoline probability and dashed lines show the boundaries of the smallest interval with at least 0.95 probability of a binomial distribution with the target probability.}
  \label{fig:  betaPlots}
\end{figure}  

Although we primarily view these isolines as a way to explore the extreme behavior of the data, it may be useful to produce visual measures of uncertainty for these estimated isolines.
To this end, we propose a block bootstrap approach, where the block size $b$ accounts for temporal dependence in the data.
For each bootstrap iteration, we obtain a resample of blocks of data $(x_{t^*,1}, x_{t^*,2}), \ldots (x_{t^*+b,1}, x_{t^*+b,2})$ where $t^*$ is randomly selected and where the total resample is of length $n$.
Then, treating this resample as data, we are able to produce an isoline as before.

Figure \ref{fig: bootstrap} shows 200 bootstrapped isolines for the Santa Ana and Karachi data, both for an exceedance probability of $p = 0.001$.
We use a block length of $b=3$ for the Santa Ana data and $b = 5$ for the Karachi data.
Shown in both plots is the original data in gray, and a single bootstrap iteration's data in black.
One notices that our simple block bootstrap technique results a subset of data with many fewer unique points, which, at a minimum, affects the GPD fit of the marginal tails and the behavior of our nonparametric method for drawing the base isoline.
Nevertheless, the bootstrapped isolines do give a useful visual representation of uncertainty.

\begin{figure}[t]
  \begin{center}
    \includegraphics[scale = .75]{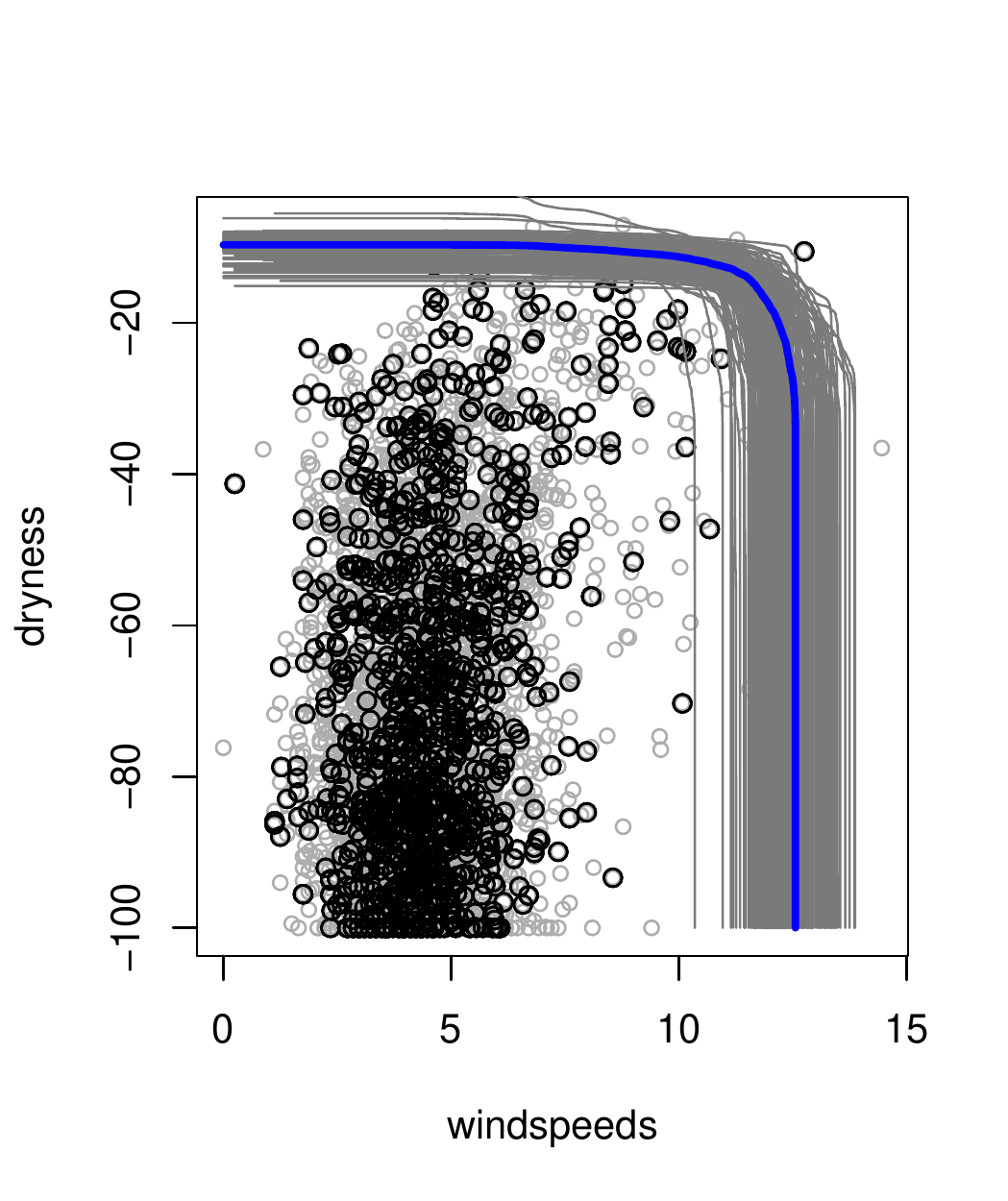}
    \includegraphics[scale = .75]{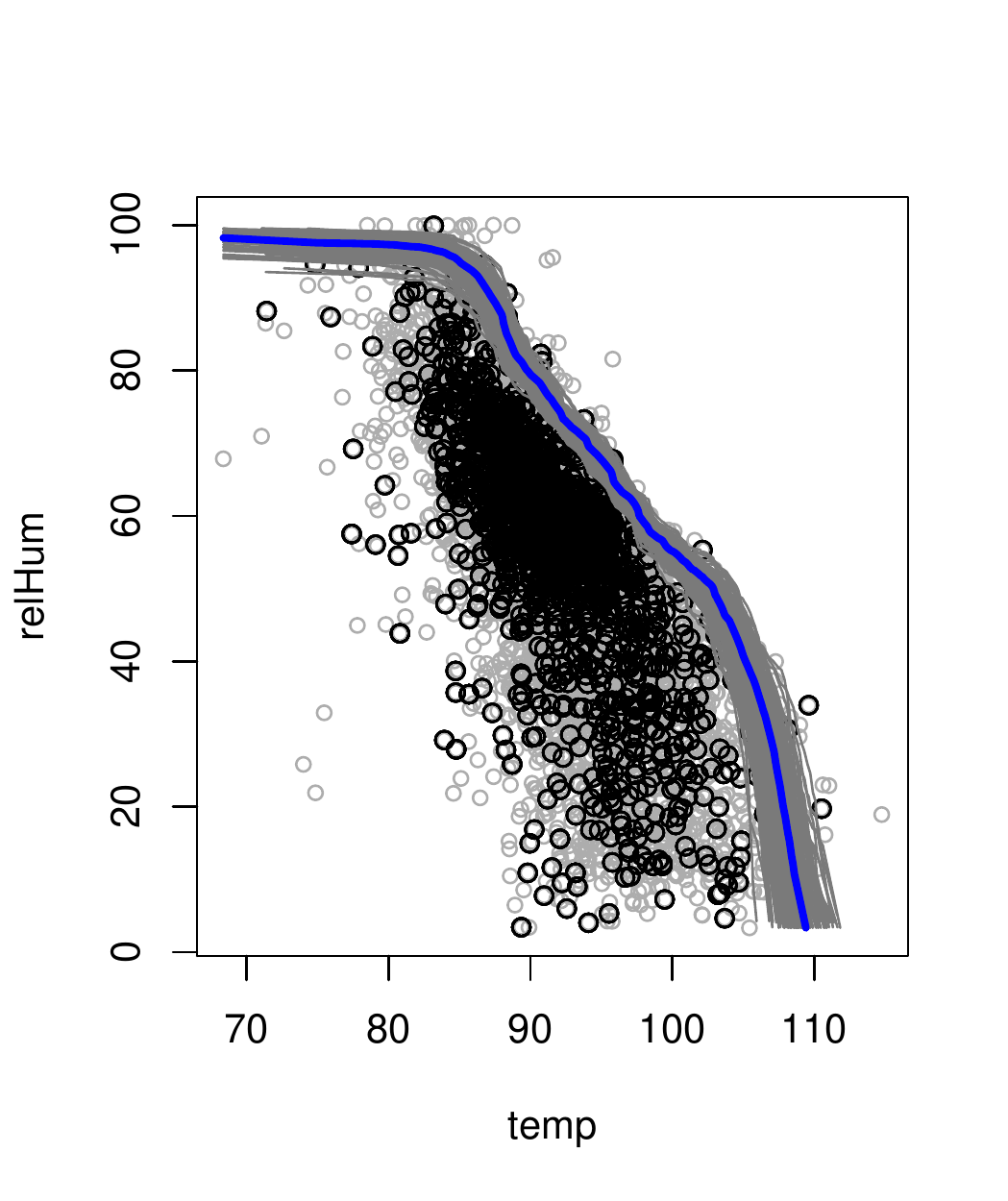}
  \end{center}
  \caption{Bootstrapped isolines for exceedance probability $p = 0.001$ for the Santa Ana (left) and Karachi data (right).  Original data shown in gray, and a single block-bootstrapped resample shown in black.}
  \label{fig:  bootstrap}
\end{figure}

%%%%%%%%%%%%%%%%%%%%%%%%%%%%%%%%%%%%%%%%%%%%%%%%%%
%CONCLUSIONS/DISCUSSION	
%%%%%%%%%%%%%%%%%%%%%%%%%%%%%%%%%%%%%%%%%%%%%%%%%%
\section{Conclusions and Discussion}
We have developed a method for producing isolines of bivariate exceedance probabilities.
This method relies on a dependence structure specifically suited for describing extremal dependence and which can be used to produce isolines for which there are few or even no exceedances.
Two advantages of the proposed method are that it is largely nonparametric, and that it can be extended to the asymptotic independent case.

This general approach is not necessarily limited to two dimensions.
Regular variation is well-developed for higher dimensions, and hidden regular variation has been described in higher dimensions as well.
However, implementing our approach in higher dimensions when the data exhibit hidden regular variation could be more complicated; for instance, different two-dimensional marginals could have different coefficients of tail dependence. 
More practically, higher-dimensional isolines would be difficult to visualize.

We believe this tool could be useful for researchers to explore the extremal behavior of bivariate data to better understand potential risk.
It is important to keep in mind however that the isolines denote the rarity of events, not impact.  
There are observations in the Karachi data which have equally low probability of exceedance to those corresponding to the June 2015 heat wave, but which likely have no human health impact.  
Hence, practical application of the technique presented here also requires additional information about which portion of the multi-variate space is impactful. 

The code and data for this project are currently posted at http://www.stat.colostate.edu/~cooleyd/Isolines/.
We are working with the maintainer of an existing {\tt R} package for extremes to implement the method within this package.

{\singlespacing
Acknowledgements:  Cooley, Thibaud, and Castillo received support from the project ``EaSM 2: Advancing extreme value analysis of high impact climate and weather events" NSF-DMS-1243102.  Wehner's contributions to this work are supported by the Regional and Global Climate Modeling Program of the Office of Biological and Environmental Research in the Department of Energy Office of Science under contract number DE-AC02-05CH11231. 
}

%%%%%%%%%%%%%%%%%%%%%%%%%%%%%%%%%%%%%%%%%%%%%%%%%%
%APPENDIX
%%%%%%%%%%%%%%%%%%%%%%%%%%%%%%%%%%%%%%%%%%%%%%%%%%
\begin{appendix}
\section{Appendix}

%1.  Start with assumption of decreasing in X.
%2.  Show isoline has negative slope.
%3.  Note that isoline in Z has negative slope
%4.  Show scaling preserves negativity.
%5.  Transformation back.  

We show that the smoothed scaling procedure in Section \ref{sec: asyIndepCase} preserves the requirement that isolines of exceedance probabilities must have negative slopes.

Assume the estimated survival function is strictly decreasing, i.e., if $\bm x_1 = (x_{1,1}, x_{1,2}) \neq \bm x_2 = (x_{2,1}, x_{2,2})$, $x_{1,1} \leq x_{2,1}$, and $x_{1,2} \leq x_{2,2}$, then $\hat {\bar F}_{\bm X}(\bm x_1) > \hat{\bar F}_{\bm X}(\bm x_2)$.
We first show that it follows that any isoline $\hat \ell_{\bm X}(p)$ must have a negative slope.
Let $\bm x_1$ and $\bm x_2$ be two distinct locations on $\hat \ell_{\bm X}(p)$.
WLOG, assume $x_{2,2} - x_{1,2} \geq 0$ and $x_{2,1} - x_{1,1} \geq 0$, implying the slope is not negative.  
This implies  $x_{2,2} \geq  x_{1,2}$ and $x_{2,1} \geq x_{1,1}$, but $\hat {\bar F}_{\bm X}(\bm x_1) = \hat {\bar F}_{\bm X}(\bm x_2)$, which is a contradiction.

As the transformation to Fr\'echet scale is monotonic, $\hat \ell_{\bm Z}(p)$ has negative slopes.

Let $\bm z_1^{(proj)}, \bm z_2^{(proj)}$ be any two points in $\hat \ell_{\bm Z}(p_{proj})$.
Let $s = p_{base}/p_{proj}$.
Let $\bm z_1^{(base)}, \bm z_2^{(base)}$ be the points in $\hat \ell_{\bm Z}(p_{base})$ such that $\bm z_i^{(proj)} = (s^{\eta_1(\bm z_i^{(base)})} z^{(base)}_{i,1}, s^{\eta_2(\bm z_i^{(base)})} z^{(base)}_{i,2})$ for $ i = 1,2$.
WLOG assume $z^{(base)}_{2,2} < z^{(base)}_{1,2}$ and  $z^{(base)}_{2,1} > z^{(base)}_{1,1}$.

Note that since $z^{(base)}_{2,1} > z^{(base)}_{1,1}$, $\eta_1(\bm z^{(base)}_2) > \eta_1(\bm z^{(base)}_1)$, and likewise since $z^{(base)}_{2,2} <  z^{(base)}_{1,2}$, $\eta_2(\bm z^{(base)}_2) < \eta_2(\bm z^{(base)}_1)$.
Hence,
$$
  z^{(proj)}_{2,2} - z^{(proj)}_{1,2} = s^{\eta_2(\bm z_2^{(base)})} z^{(base)}_{2,2} - s^{\eta_2(\bm z_1^{(base)})} z^{(base)}_{1,2} < s^{\eta_2(\bm z_1^{(base)})} ( z^{(base)}_{2,2} - z^{(base)}_{1,2}) < 0,
$$
and 
$$
  z^{(proj)}_{2,1} - z^{(proj)}_{1,1} = s^{\eta_1(\bm z_2^{(base)})} z^{(base)}_{2,1} - s^{\eta_1(\bm z_1^{(base)})} z^{(base)}_{1,1} > s^{\eta_1(\bm z_1^{(base)})} ( z^{(base)}_{2,1} - z^{(base)}_{1,1}) > 0.
$$
Thus, the slope between any two points on $\hat \ell_{\bm Z}(p_{proj})$ is negative, and since the marginal transformation is monotonic, the slope between any two points on $\hat \ell_{\bm X}(p_{proj})$ is negative.

\end{appendix}

\bibliographystyle{apalike}
\bibliography{contours.bib}

\end{document}